\documentclass[lettersize,journal]{IEEEtran}
\usepackage{amsmath,amsfonts}
\usepackage{algorithmic}
\usepackage{algorithm}
\usepackage{array}
\usepackage[caption=false,font=normalsize,labelfont=sf,textfont=sf]{subfig}
\usepackage{textcomp}
\usepackage{stfloats}
\usepackage{url}
\usepackage{verbatim}
\usepackage{graphicx}
\usepackage{cite}
\usepackage{soul}
\usepackage[dvipsnames]{xcolor}
\usepackage{multirow}
\usepackage{booktabs}

\begin{document}

\title{Multi-Modal Intelligent Channel Modeling: 
From Fine-tuned LLMs to Pre-trained Foundation Models‌}

\author{Lu~Bai,~\IEEEmembership{Senior~Member,~IEEE}, Zengrui~Han,~\IEEEmembership{Graduate~Student~Member,~IEEE}, 
Mingran~Sun,~\IEEEmembership{Graduate~Student~Member,~IEEE}, and Xiang~Cheng,~\IEEEmembership{Fellow,~IEEE}

\thanks{L.~Bai is with the Joint SDU-NTU Centre for Artificial Intelligence Research (C-FAIR), Shandong University, Jinan, 250101, P. R. China (e-mail: lubai@sdu.edu.cn).}
\thanks{Z. Han, M. Sun, and X.~Cheng are with the State Key Laboratory of Advanced Optical Communication Systems and Networks, School of Electronics, Peking University, Beijing, 100871, P. R. China (email: zengruihan@stu.pku.edu.cn, mingransun@stu.pku.edu.cn, xiangcheng@pku.edu.cn).}
}



\maketitle

\begin{abstract}
To meet the evolving demands of sixth-generation (6G) wireless channel modeling, such as precise prediction capability, extension capabilities, and system participation capability, multi-modal intelligent channel modeling (MMICM) has been proposed based on Synesthesia
of Machines (SoM) which explores the mapping relationship between multi-modal sensing in physical environment and channel characteristics in electromagnetic space. 
Furthermore, for integrating heterogeneous sensing, reasoning across scales, and generalizing to complex air-space-ground-sea communication environments, two new paradigms of MMICM are explored, including fine-tuned large language models (LLMs) for Channel Modeling (LLM4CM) and Wireless Channel Foundation Model (WiCo). 
LLM4CM leverages pre-trained LLMs on channel representations for cross-modal alignment and lightweight adaptation, enabling flexible channel modeling for 6G multi-band and multi-scenario communication systems. WiCo, which pre-trained on physically valid channel realizations and their associated environmental and modal observations, embeds electromagnetic equations for physical interpretability and uses parameterized adapters for scalability.
This article details the architectures and features of LLM4CM and WiCo, laying a foundation for artificial intelligence (AI)-native 6G wireless communication systems. Then, we conducts a comparative analysis of the two emerging paradigms, focusing on their distinct characteristics, relative advantages, inherent limitations, and performance attributes. Finally, we discuss the future research directions.

\end{abstract}

\begin{IEEEkeywords}
Multi-modal intelligent channel modeling, multi-modal sensing, fine-tuned LLMs for channel modeling, wireless channel foundation model, Synesthesia of Machines.

\end{IEEEkeywords}

\section{Introduction}
Wireless communication systems have undergone unprecedented evolution over the past few decades, advancing from the basic voice transmission of second-generation (2G) to the high-speed, low-latency connectivity of fifth-generation (5G). 
As the industry now sets its sights on sixth-generation (6G), the demands for communication performance have escalated to unprecedented levels-encompassing ultra-reliable low-latency transmission (URLLC), immersive extended reality (XR) communication, sensing-communication intelligent integrated networked applications, and ubiquitous air-space-ground-sea integrated connectivity \cite{6G}. 
Wireless channel modeling, which is a critical tool that characterizes the propagation of radio signals through these complex environments, plays a significant role in the design, optimization, and performance evaluation of 6G communication protocols, network architectures, and terminal devices. 
Therefore, the accuracy, scalability, and adaptability of channel models become pivotal to unlocking the full potential of next-generation wireless networks \cite{modeling}.

From the COST 207 in the 2G era to 3GPP TR 38.901 model for 5G, the research on standardized channel models has progressively enhanced accuracy, expanded frequency coverage, and refined scenario classification, which established a robust theoretical and empirical foundation for the design of conventional wireless communication systems. 
However, as 6G moves toward artificial intelligence (AI)-native architectures, these standardized models face inherent limitations. They cannot capture the spatio-temporal complexity of high-dynamic environments or the diversity of data required to train and validate AI-based communication systems \cite{yangmi}. To address this gap, AI-based channel modeling has emerged as a promising alternative. 
Early AI-based channel modeling, such as radio-frequency (RF)-only methods and global environment-based methods, lacks explicit representations of environmental features or cannot capture
fine-grained local environmental details around transceivers. 
To further enhance modeling accuracy for high dynamic characteristics and ultra-reliable low-latency communication requirements, multi-modal intelligent channel modeling (MMICM) \cite{mmicm}, which is inspired by Synesthesia of Machines (SoM) \cite{som}, leverages multi-modal sensing capabilities integrated with transceivers (e.g., millimeter-wave (mmWave) radar, light detection and ranging (LiDAR), red-green-blue (RGB) imaging, and depth mapping) to explore the mapping relationship between physical environments and electromagnetic spaces. For MMICM, conventional deep learning (DL) models remain constrained by poor cross-scenario generalization, inadequate multi-modal fusion, and lack of unified frameworks, which makes them unsuitable for the ultra-heterogeneous and dynamic scenarios of 6G. 

The recent breakthrough of large language models (LLMs) and pre-trained foundation models has opened new avenues for overcoming these challenges. Fine-tuned LLMs for channel modeling (LLM4CM) leverage transferable knowledge from pre-trained linguistic models to unify heterogeneous sensing and communication data, enabling coherent cross-modal fusion and lightweight adaptation to new scenarios. Meanwhile, the Wireless Channel Foundation Model (WiCo), a domain-specific pre-trained model, embeds physical propagation principles into data-driven learning, ensuring both accuracy and interpretability while supporting scalable expansion across frequencies and environments. These two paradigms represent a paradigm shift from traditional DL-based modeling, offering the potential to meet 6G’s dual demands of physical consistency and intelligent generalization.

This paper investigates two Paradigm for MMICM, LLM4CM and WiCo, including their frameworks, features, distinct characteristics, relative advantages, inherent limitations, and performance attributes. Section II reviews the development of standardized models and conventional AI-based channel modeling, and discussed the corresponding challenges and opportunities. Sections III and IV detail the frameworks and features of LLM4CM and WiCo, respectively. Section V conducts a comparative analysis of the two emerging paradigms. Finally, promising future research
directions are discussed.

\section{Channel Modeling for Future Networks}

\subsection{Standardized Channel Modeling Evolution}

In the 2G era, the COST 207 model characterized path loss in the frequency band between 150 MHz and 2 GHz, considering the distance, antenna height, terrain, and environmental effects such as shadowing and scattering. Furthermore, the COST 207 model adopted a narrowband Geometry-Based Stochastic Model (NGSM) built on extensive measurements and statistical analysis, capturing only stationary small-scale fading. With third-generation (3G), the COST 259 model characterized path loss in the frequency band between 150 MHz and 2 GHz by explicitly distinguishing between line-of-sight (LoS) and non-line-of-sight (NLoS) propagation conditions. Moreover, the COST 259 model extended NGSM to wider bandwidths and introduced multiple-input multiple-output (MIMO) support, enabling spatial correlation modeling. 
In fourth-generation (4G), the WINNER II model provided a fine-grained classification of propagation scenarios and a more accurate description of the path loss and shadowing over the 2--6 GHz frequency band. Additionally, the SCM evolved NGSM into a Geometry-Based Stochastic Model (GBSM), while the WINNER II model further incorporated dynamic scenarios to describe time-varying non-stationarity. 
In 5G, the 3GPP TR 38.901 model adopted a more refined scenario classification and extended path loss modeling to a wider frequency range of 0.5--100 GHz. Furthermore, the COST 2100 model characterized non-stationary propagation below 6 GHz through the introduction of visibility regions.
Building on the COST and WINNER frameworks, 5G channel models such as METIS, mmMAGIC, 5GCM, 3GPP TR 38.901, and IMT-2020 extended these concepts to unified multi-frequency and multi-scenario modeling.

{\color{black}In summary, the research on large-scale fading in standardized channel models has progressively advanced toward higher accuracy, finer-grained scenario classification, and broader frequency-band coverage. Meanwhile, the research on small-scale fading in standardized channel models has undergone a continuous evolution from NGSM to GBSM, from narrowband to wideband, and from stationary to non-stationary characteristics, thereby establishing a solid foundation for future channel modeling in increasingly complex scenarios.}

\subsection{AI-Based Channel Modeling Methods}

{\color{black}Future 6G communications will impose stringent and diverse requirements, motivating the deep integration of AI as a core enabler for system design and optimization~\cite{som}.
However, AI-native systems critically depend on large-scale, high-fidelity channel data, which existing standardized channel models cannot adequately provide due to their limited ability to capture spatial-temporal complexity and scenario diversity.
Consequently, increasing attention has been paid to AI-based channel models for generating massive and high-quality channel datasets.}

{\color{black} For large-scale fading, existing AI-based modeling studies can be categorized into RF-based, global environment-based, and multi-modal sensing-based methods.
RF-based methods use only RF data from measurements or ray-tracing (RT) to train deep neural networks (DNNs) for path loss modeling, but their lack of explicit environmental representation limits performance.
Global environment-based methods incorporate map data and deep learning architectures such as ResNet and VGG-16 to model propagation environments, yet their inability to capture fine-grained local details reduces reliability in highly dynamic scenarios.
To address this limitation, multi-modal sensing-based methods leverage heterogeneous sensing data, including red-green-blue (RGB) images, depth maps, and light detection and ranging (LiDAR), enabling more accurate path loss generation in dynamic environments.
For small-scale fading, most studies employ AI models trained solely on RF data, using convolutional neural networks (CNNs) and generative adversarial networks (GANs) to generate parameters such as complex amplitude, delay, angle of arrival (AoA), and angle of departure (AoD).
However, the lack of explicit environmental and geometric awareness limits their generalization and physical interpretability.
Inspired by SoM~\cite{som}, the MMICM framework~\cite{mmicm} for 6G-enabled networked intelligent systems has been proposed, motivating recent efforts to integrate RGB, LiDAR, and radar sensing with communication data for environment-aware and spatially consistent channel data generation.}

{\color{black}Conventional deep learning models for channel modeling suffer from limited generalization, as they are often trained on scenario-specific datasets and struggle to adapt to complex and dynamic environments.
Moreover, their multi-modal fusion capability is insufficient to effectively capture cross-modal correlations, hindering unified modeling.
Furthermore, the lack of a unified framework further results in task-specific and isolated designs, restricting systematic channel characteristic prediction.}

\subsection{Challenges and Opportunities: From Fine-tuned LLMs to Pre-trained Foundation Models}

With the advent of 6G, the above limitations will become increasingly critical.
6G communication systems require channel models that can simultaneously deliver high accuracy, physical interpretability, and scalability across diverse frequencies, altitudes, and environments.
They must also integrate heterogeneous data sources, including geographical information, environmental context, and user mobility, to enable comprehensive understanding and prediction of channel behavior.
Fortunately, rapid advances in AI, especially the emergence of LLMs and foundation models, have opened new opportunities for next-generation channel models with stronger cross-scenario generalization and intelligent data generation capabilities.

Adapting LLMs for channel modeling has emerged as a new paradigm for wireless channel modeling by leveraging transferable knowledge in pretrained LLMs to unify heterogeneous sensing and communication channel data in a data-driven manner. Compared with conventional deep learning models, fine-tuned LLMs exhibit three notable advantages \cite{LLM-survey}. First, cross-modal alignment enabled by natural language pretraining supports coherent fusion of modalities such as RGB-depth (RGB-D) images, LiDAR, and RF data. Second, enhanced transferability and generalization help maintain modeling accuracy across diverse scenarios and frequency bands with lightweight fine-tuning rather than retraining. Third, inherent scalability makes LLMs well suited to massive datasets, aligning with the data volume envisioned for 6G communications. Nevertheless, fine-tuned LLMs also face critical limitations. Limited domain knowledge from general-purpose corpora may hinder accurate modeling of electromagnetic propagation mechanisms. Modality extensibility remains constrained, since adding sensing modalities often requires redesigning adapters and incurs alignment overhead. Finally, dynamic adaptation is challenging, as static fine-tuned models may struggle with real-time optimization in ultra-heterogeneous 6G scenarios involving uncrewed aerial vehicles (UAVs) and rapidly time-varying propagation environments.



Recently, WiCo has been proposed as a foundation model~\cite{FM-survey} designed for multi-modal sensing and communication integration. Pretrained on massive communication and multi-modal sensing data, WiCo supports various channel data generation tasks with strong generalization and scalability.
The advantages of WiCo are mainly reflected in three aspects.
First, it embeds differentiable electromagnetic equations to jointly optimize data-driven learning and physical principles.
Second, it supports multi-modal inputs such as RF signals, geometric environment information, and textual descriptions to build physically consistent representations.
Finally, it introduces parameterized adapters to expand across different frequency bands and scenarios, reducing fine-tuning and retraining costs.

Overall, fine-tuned LLMs and WiCo provide two complementary solutions for channel modeling.
LLM4CM leverages transferable cross-modal priors and lightweight adaptation, making it suitable for resource-constrained or task-specific applications.
In contrast, WiCo is pretrained on channel-centric and physically grounded corpora, mitigating the limited propagation awareness and modality extensibility of general-purpose LLMs.

\section{Fine-tuned LLM for Wireless Channel}

\begin{figure*}[t]
\centering
\includegraphics[width=\textwidth]{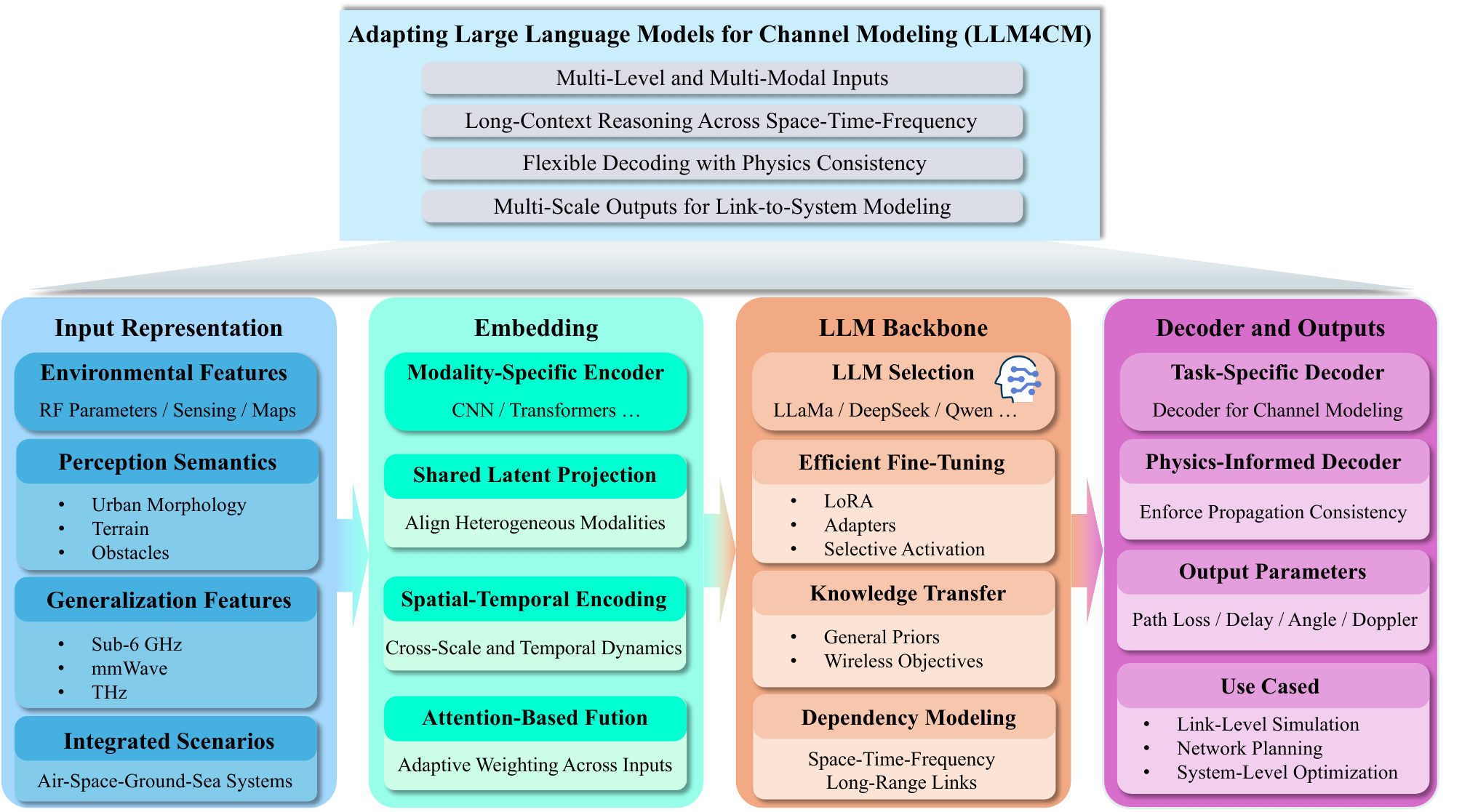}
\caption{Holistic Design Pipeline of LLM4CM: Input, Embedding, LLM Backbone, and Output.
}
\label{llm4cm}
\end{figure*}

LLMs, originally developed for natural language processing, have recently shown strong potential as general-purpose foundation models capable of learning complex structures across diverse modalities.
In wireless communications, fine-tuned LLMs provide a new paradigm for channel modeling by unifying heterogeneous data, capturing long-range dependencies, and generalizing across environments and deployment scenarios.
Unlike conventional stochastic and deep learning–based channel models, LLM-based approaches emphasize data-driven representation learning and cross-domain knowledge transfer.
By fine-tuning LLMs with wireless-specific inputs and objectives, scalable and flexible system-level channel models can be constructed, adapting to diverse propagation conditions, frequency bands, and network architectures, thereby enabling flexible channel modeling for 6G multi-band and multi-modal systems and supporting wideband spectrum utilization and sensing–communication integration.
This section introduces the LLM4CM paradigm, illustrated in Fig.~\ref{llm4cm}, including the input representation, embedding module, LLM backbone, decoder, and output representation.

\subsection{Input Representation}
The input of LLM4CM incorporates multi-level and multi-modal features describing the wireless propagation environment, which can be categorized into three dimensions.
First, environmental features characterize physical and electromagnetic properties, including RF parameters, geometric relationships, and spatial context from global or local maps.
Second, perception-oriented features capture scene-level semantics such as urban morphology, terrain type, building distribution, traffic intensity, and dynamic obstacles, typically obtained from vision sensors, LiDAR, or remote sensing data.
Third, generalization-oriented features enhance cross-scenario robustness through multi-band fusion across sub-6 GHz, mmWave, and Terahertz (THz) bands, as well as cross-domain abstractions spanning terrestrial, aerial, maritime, and satellite communications.
Rather than targeting a specific application, LLM4CM is designed for air–space–ground–sea integrated communication systems, requiring inputs that are both physically grounded and semantically rich to enable large-scale generalization.

\subsection{Embedding Module}
The embedding module performs feature extraction, alignment, and fusion across heterogeneous inputs.
Structured numerical features and unstructured data, such as images, maps, and point clouds, are first encoded by modality-specific networks (e.g., convolutional or transformer-based models) and then projected into a unified latent space compatible with the LLM backbone, enabling cross-modal alignment and joint reasoning.

At this stage, feature fusion strategies~\cite{ff} play a key role in aggregating heterogeneous representations by adaptively emphasizing task-relevant information across modalities, frequencies, spatial scales, and temporal dynamics.
As a result, the fused embeddings provide a high-level abstraction of the physical environment, effectively bridging multi-modal sensing information and the natural language domain.

\subsection{LLM Backbone Module}
The LLM backbone serves as the core reasoning and representation engine, where partial fine-tuning strategies, such as adapter layers, low-rank adaptation, and selective attention activation~\cite{fine-tuning}, are commonly adopted to balance performance and efficiency.

Various architectures, including dense Transformers and mixture-of-experts (MoE) variants~\cite{MoE}, can be utilized, with open-source LLMs (e.g., LLaMA, DeepSeek, et al.) offering transparency and flexibility for domain-specific fine-tuning, as summarized in Table~\ref{LLM}.
Owing to their long-context modeling and unified latent representations, LLMs can capture long-range dependencies across spatial, temporal, and frequency domains, outperforming conventional deep learning models for wireless channel modeling.
Moreover, by leveraging general knowledge acquired through large-scale pretraining, LLMs can encode high-level, domain-agnostic priors over environmental structure and contextual relationships, thereby exhibiting advantages over conventional deep learning models in channel modeling. However, such priors remain distinct from the domain-specific inductive biases introduced by foundation models pretrained on communication-centric data. In addition, the strong generalization capability of LLMs allows learned representations to transfer effectively across unseen environments and deployment scenarios, reducing the need for exhaustive retraining and supporting scalable channel modeling in complex 6G communications.

\begin{table*}[htbp]
\centering
\footnotesize
\renewcommand{\arraystretch}{1.25}
\caption{Comparison of Representative Large Language Models from a Channel Modeling Perspective}
\label{LLM}
\begin{tabular}{p{2.2cm} p{3.1cm} p{3.2cm} p{3.6cm} p{3.6cm}}
\hline
\textbf{Model} & \textbf{Channel-Relevant Focus} & \textbf{Architecture} & \textbf{Strengths for Channel Modeling} & \textbf{Limitations for Channel Modeling} \\
\hline

Gemma 3 &
Multimodal environment perception and contextual understanding &
Dense Transformer &
Facilitates joint modeling of visual environmental cues and contextual information, which is beneficial for environment-aware channel characterization &
Limited model scale and lack of communication-specific inductive biases may restrict modeling accuracy \\

\hline

Llama 4 &
Long-context dependency modeling and multimodal reasoning &
MoE Transformer &
Capable of capturing long-range temporal and contextual dependencies in dynamic channel environments with improved efficiency enabled by MoE architecture &
Multimodal capability is primarily vision--language oriented and not explicitly designed for wireless physical modeling \\

\hline

DeepSeek-V3.2 &
General-purpose representation learning and structured sequence modeling &
MoE Transformer &
Provides strong expressive capacity and scalable representations that can be adapted to complex channel modeling tasks under diverse scenarios &
Primarily pretrained on text-centric data, lacking explicit incorporation of wireless-domain physical priors \\

\hline

Qwen 3 &
General-purpose contextual modeling across heterogeneous tasks &
MoE Transformer &
Offers flexible representations and efficient inference mechanisms that can be transferred to data-driven channel modeling frameworks &
Reasoning depth and long-context modeling capability may be insufficient for highly time-varying or complex channel dynamics \\

\hline
\end{tabular}
\end{table*}

\subsection{Decoder Module}
The decoder module maps the high-level latent representations produced by the LLM backbone into task-specific channel outputs. This module can be instantiated with different decoder architectures depending on the target channel modeling objective.  For generation of channel fading characteristics, transformer-based decoders are more suitable due to their ability to model global dependencies and structural correlations in high-dimensional channel representations. Furthermore, physics-informed decoders can be incorporated to explicitly enforce known propagation laws, thereby ensuring physical consistency and interpretability.

\subsection{Output Representation}
The output of the framework can take multiple forms and scales. At a fine temporal scale, outputs may consist of time-series numerical values, such as path loss, delay spread, Doppler spectrum, or full channel impulse responses. At a spatial scale, outputs may be map-based representations, including coverage maps, blockage probability maps, or spatial correlation distributions. This flexibility enables the framework to support applications ranging from link-level simulation to network planning and system-level optimization, thereby meeting the diverse and evolving channel modeling requirements of 6G communication systems.

In summary, fine-tuned LLMs provide a unifying and extensible framework for next-generation wireless channel modeling, capable of integrating heterogeneous inputs, reasoning across scales, and generalizing to complex air-space-ground-sea communication environments. Beyond these functional advantages, LLM4CM represents a paradigm shift from conventional deep learning models toward LLM-based approaches that enable new forms of system-level reasoning, adaptability, and scalability. Therefore, such models are better positioned than conventional channel modeling techniques to address the stringent requirements and emerging challenges of 6G communication systems, including ultra-heterogeneity, dynamic deployment, and large-scale intelligent networking, et al.

\section{Pre-training Foundation Model for Wireless Channel}

\begin{figure*}[t]
\centering
\includegraphics[width=\textwidth]{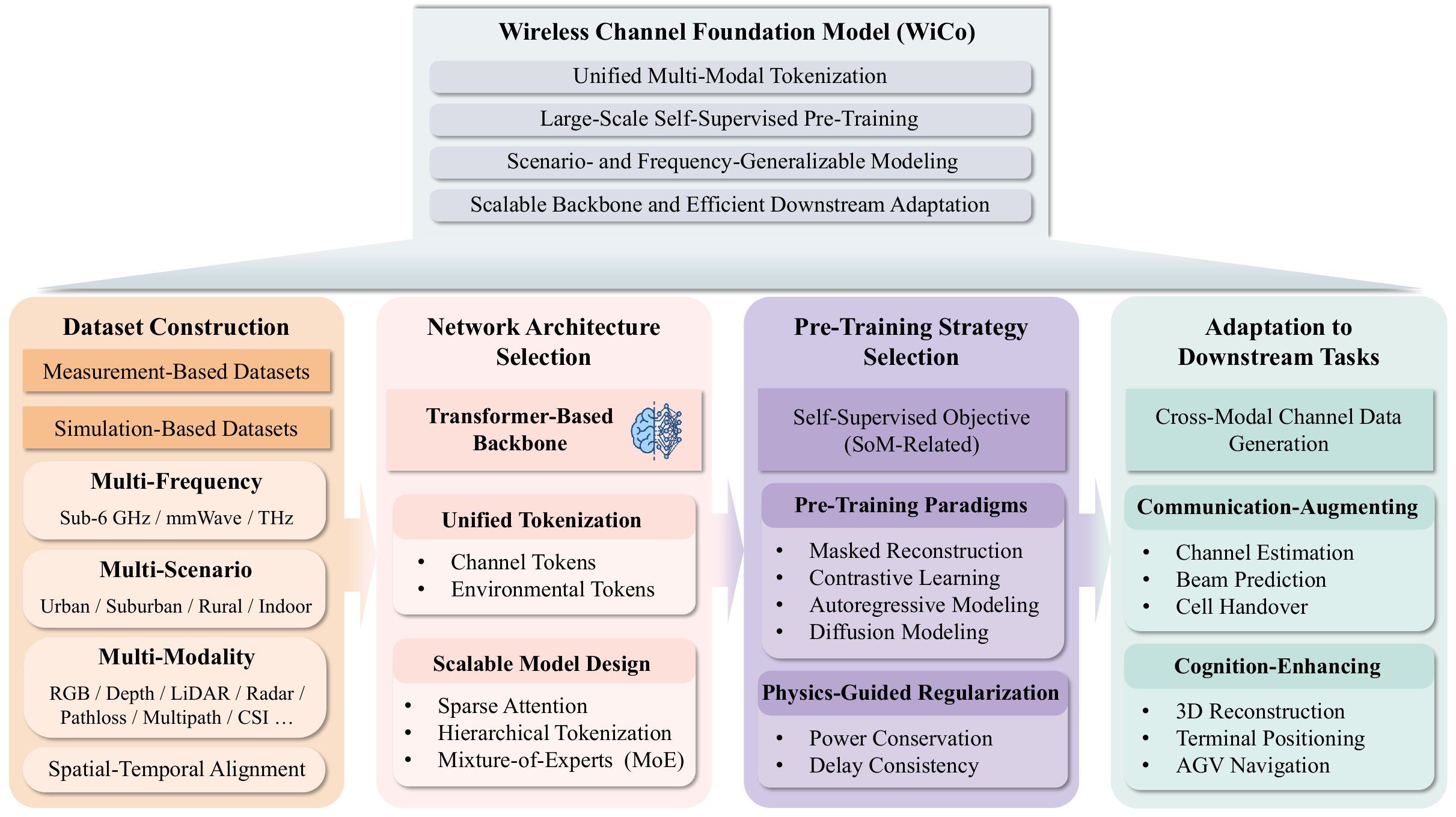}
\caption{Holistic Design Pipeline of WiCo: Dataset, Architecture, Pre-training, and Adaptation.
}
\label{FM}
\end{figure*}

With the continued development of AI-native communication systems and cross-modal learning methodologies, 
WiCo is pre-trained from scratch as a domain-specific foundation model intended for such communication systems.
Its primary role is the cross-modal generation of large-scale, high-quality channel data from heterogeneous sensing inputs (e.g., maps, LiDAR, depth/RGB images, and network telemetry), thereby enabling the training and evaluation of models designed for various communication tasks under diverse channel conditions.
Unlike fine-tuned LLMs, which adapt linguistic priors to new tasks, WiCo is pre-trained directly on channel-centric corpora so that physically grounded representations are acquired prior to any task adaptation.
To establish WiCo for AI-native communication systems, its design must be considered from a holistic perspective encompassing dataset construction, architectural scalability, pre-training methodology~\cite{pretrain}, and downstream adaptability.
In this section, WiCo is introduced as show in Fig.~\ref{FM}, including dataset requirements, network architecture selection, pre-training strategy selection, and adaptation to downstream tasks.

\subsection{Dataset Requirements}



The performance of WiCo largely depends on the scale and diversity of its pre-training dataset. To support large-scale and task-agnostic pre-training, the dataset should contain heterogeneous samples covering multiple frequency bands, such as sub-6 GHz, mmWave, and THz, diverse propagation environments, including indoor, urban, and suburban scenarios, as well as multiple modalities, such as channel matrices, LiDAR point clouds, digital maps, and other related representations. All modalities need to be temporally and spatially aligned to preserve physical consistency between the propagation environment and the corresponding channel response.

Two main data acquisition methods can be employed \cite{synthsom}. Measurement-based datasets provide high-fidelity multi-modal data that capture realistic propagation and environmental characteristics, while simulation-based approaches enable large-scale and controllable generation of channels and associated environmental modalities under diverse configurations. Either source can be used independently for pre-training, whereas their combination further improves realism and diversity. Through such data construction, WiCo can learn generalizable and physically consistent representations that support large-scale pre-training and cross-modal channel generation.

\subsection{Network Architecture Selection}



The architecture of WiCo is designed to ensure scalability, universality, and cross-modal adaptability. Transformer-based~\cite{transformer} structures are adopted as the backbone, as self-attention enables efficient modeling of long-range spatial-temporal dependencies in wireless channels. This also supports flexible integration of heterogeneous modalities, such as environmental features and propagation parameters, within a unified latent representation space.

To further improve efficiency and capacity, hierarchical tokenization and sparse attention can capture fine-grained local variations while maintaining global consistency. The feed-forward layers may be augmented by mixture-of-experts modules to improve efficiency and support scalable expansion without retraining the entire network. Such a design follows the extensibility principle of foundation models, where increasing model depth or width leads to stronger generalization.

Through this architecture, WiCo can represent multi-modal information in a unified latent space and generate channel responses conditioned on complex environmental contexts. 

\subsection{Pre-training Strategy Selection}
WiCo is pre-trained in a self-supervised manner to acquire general and physically consistent representations of wireless channels. Since labeled data are often limited, the model is trained on heterogeneous datasets using task-agnostic but SoM-related objectives to capture intrinsic propagation patterns and cross-modal correlations. The training data are restricted to physically valid channel realizations with corresponding environmental and modal observations.

WiCo adopts three complementary pre-training paradigms~\cite{pretrain}. Masked reconstruction is used to recover missing channel tokens from contextual information, with a focus on learning spatial-temporal continuity and structural sparsity in channel representations. 
Contrastive learning emphasizes cross-modal alignment by associating channel representations with corresponding environmental features, thereby capturing consistent semantic and physical relationships across modalities. 
In addition, autoregressive and diffusion modeling focus on the dynamic and stochastic properties of wireless channels, enabling the model to characterize temporal evolution.
Together, these pre-training paradigms provide a unified learning framework that captures structural regularities, cross-modal semantics, and dynamic evolution of wireless channels, laying a foundation for robust and generalizable channel modeling.

To enable joint optimization between data-driven learning and physical principles, physics-guided regularizations are incorporated by embedding differentiable constraints, such as power conservation, bandwidth-delay duality, and angular wrapping. Through large-scale self-supervised pre-training, WiCo develops invariant and scalable representations across frequencies, scenarios, and antenna configurations, enabling emergent reasoning and efficient downstream adaptation.

\subsection{Adaptation to Downstream Tasks}
After pre-training, WiCo can be efficiently adapted to various downstream tasks, with the primary objective being cross-modal channel data generation under diverse scenarios and communication conditions.
Through lightweight fine-tuning, the model can be quickly adapted to new frequency bands, antenna configurations, and propagation environments at low computational cost.
The learned representations capture both environmental semantics and physical propagation features, which can further support perception-assisted communication tasks such as beam selection, blockage prediction, and localization.
Such adaptation exploits the strong generalization ability of the pre-trained foundation model, enabling WiCo to generate physically consistent, high-quality channel data and to extend its utility across heterogeneous sensing modalities and dynamic communication contexts.

Overall, WiCo functions as a reusable and extensible foundation model for AI-native 6G communication systems. 
It unifies heterogeneous sensing modalities within a shared latent representation space, enables scalable and physically consistent channel generation, and supports efficient adaptation to a wide range of downstream tasks with minimal supervision. 
By transferring domain-specific priors learned through large-scale self-supervised pre-training, WiCo bridges environmental sensing and wireless channel, improving robustness and generalization across diverse channel conditions.

\section{Comparison and Discussion}

This section compares and discusses the two emerging paradigms for MMICM. 
Although both paradigms aim to enable cross-modal channel data generation, they differ substantially in capability formation, generalization behavior, and data generation quality.
We first present a systematic comparison to highlight their respective strengths and limitations, followed by representative case studies that illustrate their performance in practical channel modeling scenarios. 
Based on these observations, we further discuss open challenges and potential future research directions for MMICM in AI-native 6G communication systems.

\subsection{Comparison Between the Two Roadmaps}

MMICM have gradually converged toward two representative paradigms, i.e., fine-tuned LLM for wireless channel and pre-training foundation model for wireless channel. They differ in architecture design, data utilization, and generalization mechanisms, leading to distinct trade-offs in efficiency, robustness, and deployment flexibility.

The first paradigm, i.e., fine-tuned LLM for wireless channel, builds MMICM on general-purpose foundation models through knowledge transfer and task-oriented fine-tuning. It requires minimal redesign and relatively limited channel data, while heterogeneous modalities such as RF signals, images, and environmental descriptors can be integrated via lightweight adapters. This enables rapid deployment with low engineering cost, but performance is often constrained by the modality scope and semantic bias of general-purpose pretraining, making it more reliable for simpler tasks or scenarios close to the fine-tuning distribution.

In contrast, the second paradigm, i.e., pre-training foundation model for wireless channel, establishes MMICM through large-scale pretraining within the wireless domain. It explicitly accounts for electromagnetic propagation, multi-scale spatial-temporal dependencies, and cross-modal physical consistency. Trained on massive, physically grounded multi-modal datasets, it learns native wireless representations linking environments, sensing modalities, and channel responses. Although it incurs higher design and data curation costs, it offers stronger robustness under distribution shifts and better few-shot or zero-shot deployment, providing a principled basis for high-fidelity cross-modal channel generation.

In summary, both paradigms offer complementary pathways toward MMICM for future 6G networks.
A systematic comparison of the two paradigms is presented in Table~\ref{com}.
\begin{table*}[t]
\centering
\caption{Comparison between Paradigm 1 and Paradigm 2 for MMICM}
\label{com}
\renewcommand{\arraystretch}{1.25}
\setlength{\tabcolsep}{6pt}

\begin{tabular}{>{\centering\arraybackslash}p{2cm}|>{\centering\arraybackslash}p{7.5cm}|>{\centering\arraybackslash}p{7.5cm}}
\hline
& \textbf{Paradigm 1: Fine-tuned LLM for wireless channel} 
& \textbf{Paradigm 2: Pre-training foundation model for wireless channel} \\
\hline

Advantages 
& Leverages mature general-purpose foundation models to enable cross-modal channel data generation with minimal network redesign and rapid deployment 
& Learns native wireless representations through large-scale multi-modal pretraining, enabling high-fidelity and physically consistent channel data generation \\
\hline

Limitations 
& Cross-modal representations are constrained by the modality scope and semantic bias of general-purpose pretraining, limiting physical consistency in generated channel data 
& Requires large-scale heterogeneous dataset construction and wireless-specific model design to capture propagation characteristics \\
\hline

Problem-solving capability 
& Suitable for generating channel data for relatively simple scenarios or tasks close to the fine-tuning distribution 
& Capable of generating reliable channel data for complex scenarios with strong dynamics and heterogeneous system configurations \\
\hline

Generalization capability 
& Relies on task-specific fine-tuning to maintain generation quality under moderate distribution shifts 
& Achieves strong few-shot or zero-shot channel data generation performance across new scenarios and system settings \\
\hline

Task universality 
& Limited support for diverse channel generation tasks, often requiring separate adaptation for different modalities or objectives 
& Provides a unified MMICM framework that supports multiple channel generation tasks and heterogeneous modalities within a single model \\
\hline

Overhead 
& Low model design overhead but relatively high computational and storage cost during deployment 
& High pretraining and data curation cost, but reduced adaptation and maintenance overhead in downstream MMICM applications \\
\hline

\end{tabular}

\end{table*}

\subsection{Case Studies}

\subsubsection{Path Loss Generation}

To compare the performance of the two representative MMICM paradigms for path loss map generation, a UAV-to-ground communication scenario is considered. As shown in Fig. 3(a), RGB images captured by a UAV-mounted camera provide a top-down view of the urban environment, implicitly conveying information about building distribution, street layout, and blockage conditions. Based on this visual input, the model generates the path loss maps between the UAV and a grid of ground antennas, enabling massive-scale and high-quality path loss map generation for 6G communication system design and development.

\begin{figure}[t]
\centering
\includegraphics[width=0.5\textwidth]{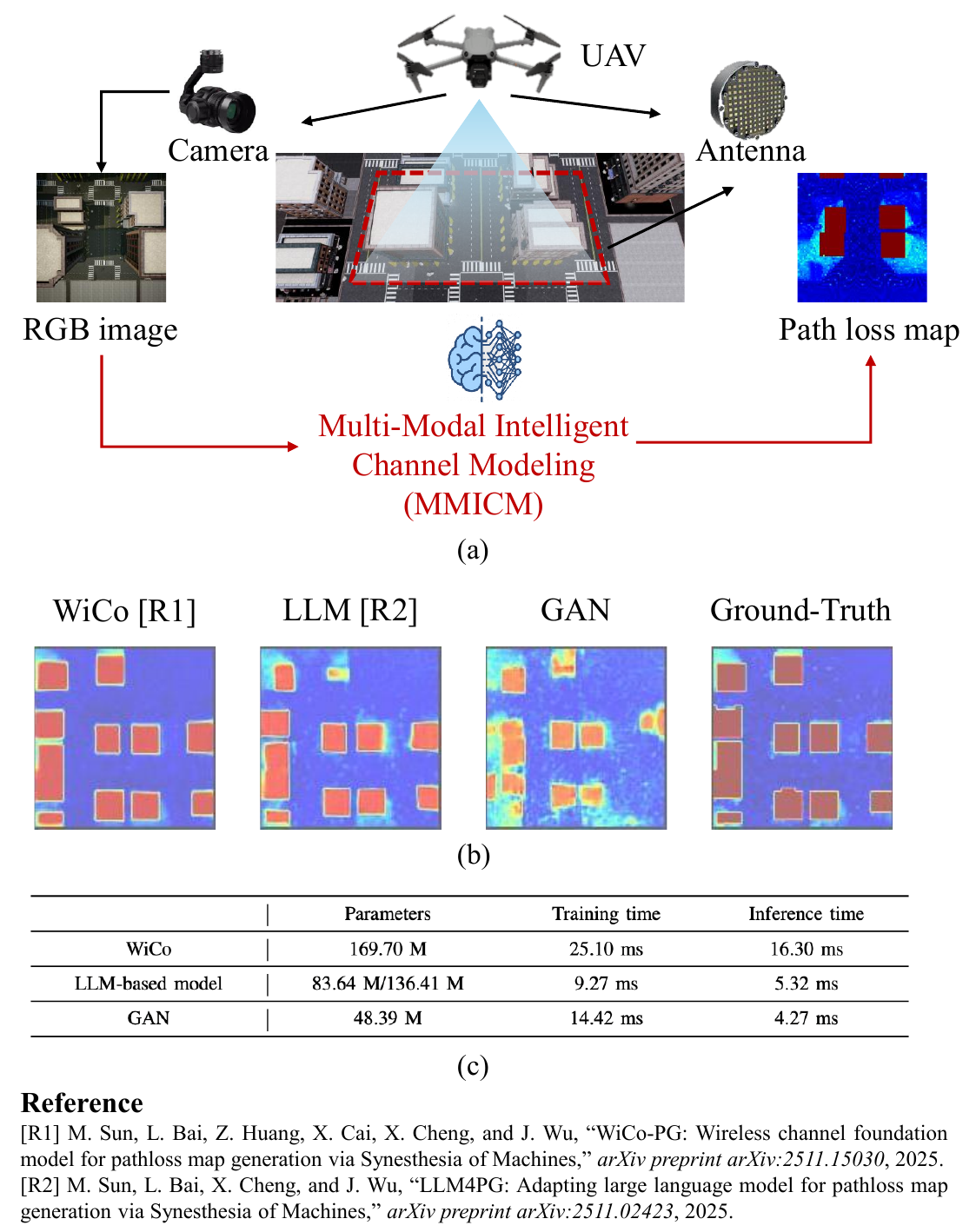}
\caption{Case study for channel pathloss map generation.
(a) UAV-captured RGB image-based generation of pathloss map over a ground antenna grid.
(b) Qualitative comparison of generated pathloss maps using different models and ray tracing.
(c) Comparison of model complexity and computational overhead.
}
\label{PL}
\end{figure}

Fig.~\ref{PL} illustrates the path loss maps generated by different approaches. Compared with LLM-based methods~\cite{GPT} and conventional deep learning models, the proposed WiCo achieves higher consistency with the ray-tracing reference, particularly in capturing sharp loss variations caused by building obstructions and shadowing effects. In contrast, the GAN-based and LLM-based models lacking wireless-domain awareness tend to produce overly smooth path loss patterns, failing to preserve the spatial discontinuities inherent in urban propagation environments. These results highlight the benefit of large-scale pretraining for accurate path loss generation from visual image inputs.

From a computational perspective, although WiCo adopts a foundation-model architecture with increased model capacity, its inference latency remains practical. Overall, WiCo strikes a favorable balance between modeling accuracy and computational efficiency, making it well suited for massive-scale path loss map generation in AI-native wireless systems.

\subsubsection{Multipath Generation}
\begin{figure}[t]
\centering
\includegraphics[width=0.5\textwidth]{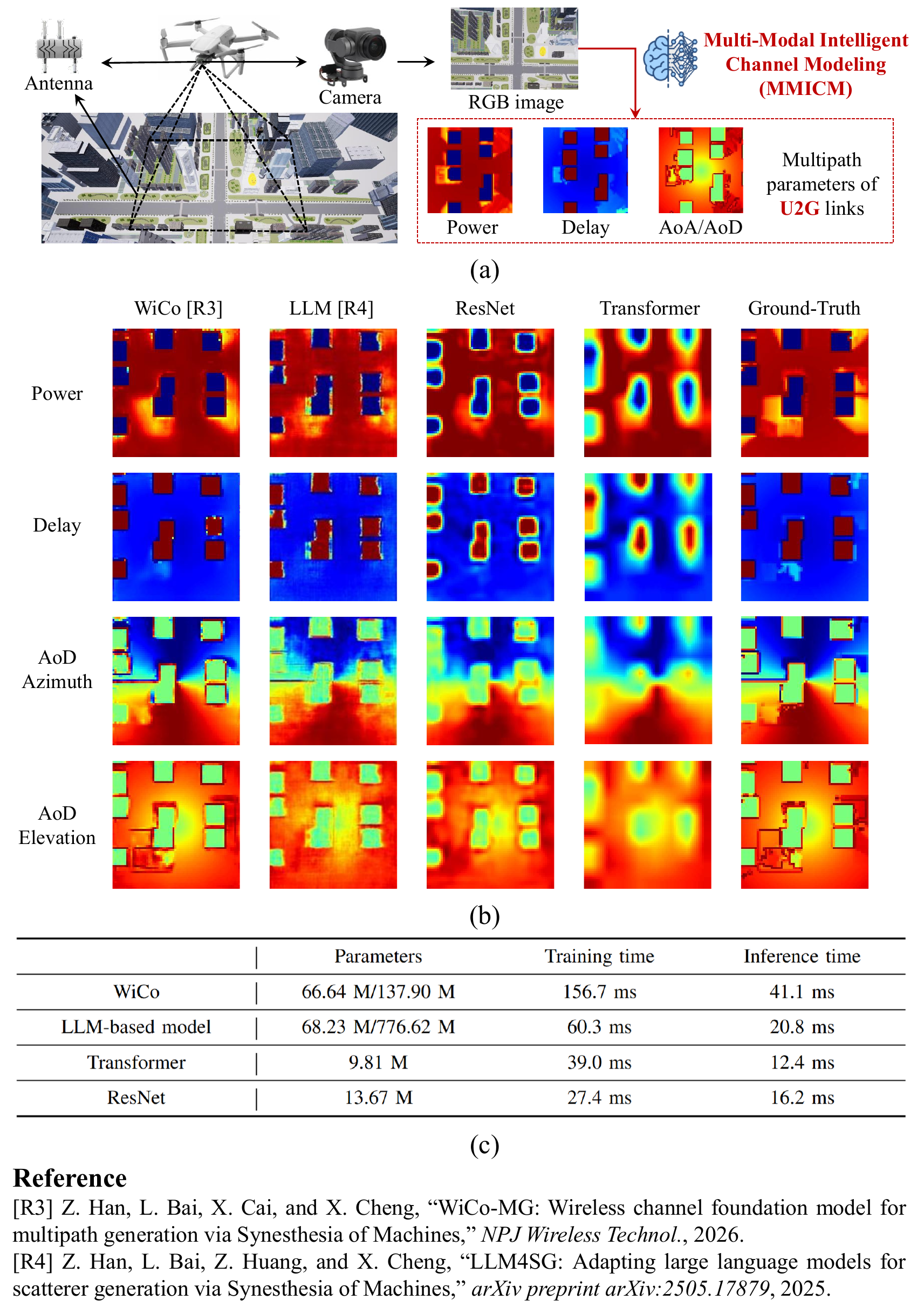}
\caption{Case study for channel multipath generation.
(a) UAV-captured RGB image-based generation of multipath channel parameters over a ground antenna grid.
(b) Qualitative comparison of generated channel parameter maps using different models and ray tracing.
(c) Comparison of model complexity and computational overhead.
}
\label{multipath}
\end{figure}
While the previous case study examines large-scale channel characteristics via path loss map generation, this case study further evaluates MMICM for small-scale fading through multipath parameter generation. Accurate multipath modeling requires learning fine-grained spatial variations and physically consistent relationships among paths, delays, and angular parameters, making this task a stricter test of cross-modal channel generation and the two MMICM paradigms.

As illustrated in Fig.~\ref{multipath}(a), RGB images from a UAV-mounted camera are used to infer channels between the UAV antenna and a grid of ground antennas. The top-down images implicitly encode scene geometry, building layout, and blockage. Conditioned on this visual input, the MMICM generates multipath parameters over the grid, including received power, propagation delay, AoD, and AoA, enabling channel generation without exhaustive measurements or ray tracing.

Fig.~\ref{multipath}(b) compares channel parameter maps generated by WiCo, an LLM-based model~\cite{GPT}, conventional models (ResNet and Transformer), and RT as reference. WiCo shows the closest consistency with RT across all parameters, especially in capturing building-induced discontinuities and preserving smooth delay and angular variations. By contrast, non-wireless-pretrained models often produce over-smoothed or distorted patterns, particularly for angular parameters, revealing limited ability to preserve multipath structure. This highlights the benefit of wireless-native pretraining for physically plausible cross-modal channel generation.

Fig.~\ref{multipath}(c) further compares complexity and overhead in parameter count, training time, and inference time. Although WiCo is larger due to its foundation-model design, it remains computationally feasible with competitive inference latency. In contrast, LLM-based models incur substantial parameter overhead, while conventional models are lighter but yield inferior quality, as shown in Fig.~\ref{multipath}(b). Overall, WiCo achieves a favorable balance between fidelity and computational cost for large-scale channel data generation in AI-native wireless systems.

\subsection{Open Issues}

As the progress of MMICM, the field is increasingly converging toward two representative paradigms, namely LLM4CM and WiCo. 
Although both aim to enable cross-modal channel data generation for AI-native 6G systems, their fundamentally different capability formation mechanisms lead to distinct limitations and research priorities. The following open issues are critical for shaping the future evolution of these paradigms.

\subsubsection{Paradigm-Specific Physical Reliability and Interpretability}

A key issue is ensuring physical reliability while preserving each paradigm's strengths. For LLM4CM, aligning general-purpose pretraining priors with electromagnetic constraints remains difficult, and physically implausible generations may occur despite cross-modal coherence. For WiCo, wireless-native corpora and physics-guided regularization improve interpretability, yet generalizing constraints across frequency bands, antenna configurations, and sensing modalities remains challenging without sacrificing flexibility. Paradigm-aware strategies balancing physical consistency and interpretability remain open.

\subsubsection{Scalability and Robust Generalization Toward Open-World 6G Scenarios}
Another issue is robust generalization under open-world and extreme deployments. LLM4CM enables efficient adaptation, but performance in highly dynamic scenarios such as integrated air-space-ground-sea networks still depends on fine-tuning coverage and modality alignment. WiCo supports few-shot or zero-shot transfer via large-scale pretraining, but scalability is constrained by massive, physically consistent, spatially aligned multi-modal datasets. How to evolve toward open-world 6G through continual learning or hybrid adaptation-pretraining remains unresolved.

\subsubsection{Unified and Paradigm-Fair Evaluation Protocols}
The lack of unified and paradigm-fair evaluation protocols is another challenge. Existing evaluations often rely on visual inspection or coarse statistical comparisons with ray tracing, failing to capture cross-modal consistency and system-level relevance. Criteria should also reflect different objectives, including adaptation efficiency, flexibility, and wireless-native fidelity. Standardized benchmarks covering these aspects are essential.

\subsubsection{From Channel Data Generation to Embodied and Closed-Loop Intelligence}
An emerging issue is how MMICM supports embodied intelligence~\cite{embodied} in sensing-communication integrated systems. 
In scenarios with UAVs, robots, or autonomous agents, channel modeling becomes part of a closed-loop perception-decision-action process coupling communication, sensing, and mobility. 
This raises challenges in interactive physical consistency, long-term temporal coherence, and uncertainty propagation to closed-loop control and safety. 
Enabling channel foundation models to interface naturally with embodied agents may be a key frontier for AI-native 6G systems.

\section{Conclusions}
This article systematically reviews wireless channel modeling's evolution and innovations. Standardized models have advanced in accuracy and frequency coverage but fail to meet the requirements of AI-native 6G wireless communication systems. Traditional AI-based models lack generalization and unified frameworks, motivating the development of LLM4CM and WiCo.
LLM4CM enables cross-modal data fusion and flexible scenario adaptation, while WiCo integrates physical principles with data-driven learning, ensuring interpretability and scalability. 
The two paradigms both address key gaps, better suiting the ultra-heterogeneity, dynamic deployment, and generalization scalability of 6G wireless communication systems.
Future research could expand developing paradigm-aware strategies for balancing physical consistency and interpretability, designing scalable robust generalization mechanisms for open-world 6G deployment, and establishing standardized, paradigm-fair evaluation protocols to advance MMICM.



\vfill


\begin{thebibliography}{29}
\bibitem{6G}
A. Fayad, T. Cinkler, and J. Rak, ``Toward 6G optical fronthaul: A survey on enabling technologies and research perspectives," \emph{IEEE Commun. Surveys Tuts.}, vol. 27, no. 1, pp. 629--666, Feb. 2025.

\bibitem{modeling}
A. Molisch, \emph{Wireless Communications}. UK: John Wiley Sons, 2011.

\bibitem{yangmi}
M. Yang \emph{et al.}, ``AI-enabled data-driven channel modeling for future communications," \emph{IEEE Commun. Mag.}, vol. 62, no. 4, pp. 112--118, Apr. 2024.

\bibitem{mmicm}
L. Bai \emph{et al.}, ``Multi-modal intelligent channel modeling: A new modeling paradigm via Synesthesia of Machines," \emph{IEEE Commun. Surveys Tuts.}, vol. 28, pp. 2612--2649, 2026.

\bibitem{som}
X. Cheng \emph{et al.}, ``Intelligent multi-modal sensing-communication integration: Synesthesia of Machines,'' \emph{IEEE Commun. Surveys Tuts.}, vol.~26, no.~1, pp.~258--301, Firstquarter 2024. 

\bibitem{LLM-survey}
S. Minaee \emph{et al.}, ``Large language models: A survey,'' \emph{arXiv preprint arXiv:2402.06196}, 2024.


\bibitem{FM-survey}
R.~Bommasani \emph{et~al.}, ``On the opportunities and risks of foundation models,'' 
\emph{arXiv preprint arXiv:2108.07258}, 2021.


\bibitem{ff}
P. P. Liang, A. Zadeh, and L. -P. Morency, ``Foundations \& trends in multimodal machine learning: Principles, challenges, and open questions,'' \emph{ACM Comput. Surv.}, vol. 56, no. 10, pp. 1--42,  Oct. 2024.

\bibitem{fine-tuning}
H. Wu, X. Chen, and K. Huang, ``Device-edge cooperative fine-tuning of foundation models as a 6G service," \emph{IEEE Wirel. Commun.}, vol.~31, no.~3, pp.~60--67, Jun.~2024.

\bibitem{MoE}
J. Wang \emph{et al.}, ``Toward scalable generative AI via mixture of experts in mobile edge networks," \emph{IEEE Wirel. Commun.}, vol.~32, no.~1, pp.~142--149, Feb. 2025.  


\bibitem{pretrain}
S. R. K. Kottapalli \emph{et al.}, ``Foundation models for time series: A survey," \emph{arXiv preprint arXiv:2504.04011}, 2025.

\bibitem{synthsom}
X. Cheng \emph{et al.}, ``SynthSoM: A synthetic intelligent multi-modal sensing-communication dataset for Synesthesia of Machines (SoM)," \emph{Sci. Data}, vol. 12, pp. 819--833, May 2025.

\bibitem{transformer}
Y. Wang \emph{et al.}, ``Transformer-empowered 6G intelligent networks: From massive MIMO processing to semantic communication," \emph{IEEE Wirel. Commun.}, vol.~30, no.~6, pp.~127--135, Dec. 2023.  






\bibitem{GPT}
Radford, A. \emph{et al.}
Language models are unsupervised multitask learners.
OpenAI Blog (2019).

\bibitem{embodied}
M. Lisondra, B. Benhabib, and G. Nejat, ``Embodied AI with foundation models for mobile service robots: A systematic review," \emph{arXiv preprint arXiv:2505.20503}, 2025.

\end{thebibliography}
\end{document}